# *In-situ* angle-resolved photoemission spectroscopy of copper-oxide thin films synthesized by molecular beam epitaxy

Chung Koo Kim[a], Ilya K. Drozdov[a], Kazuhiro Fujita[a], J. C. Séamus Davis[a,c,d,e], Ivan Božović[a,b], Tonica Valla[a]

[a.] *CMPMS Department, Brookhaven National Laboratory, Upton, NY 11973, USA*
[b.] *Applied Physics Department, Yale University, New Haven, Connecticut 06520, USA*
[c.] *LASSP, Department of Physics, Cornell University, Ithaca, NY 14853, USA*
[d.] *School of Physics and Astronomy, University of St. Andrews, Fife KY16 9SS, Scotland*
[e.] *Kavli Institute at Cornell for Nanoscale Science, Cornell University, Ithaca, NY 14853, USA*

Angle-resolved photoemission spectroscopy (ARPES) is the key momentum-resolved technique for direct probing of the electronic structure of a material. However, since it is very surface-sensitive, it has been applied to a relatively small set of complex oxides that can be easily cleaved in ultra-high vacuum. Here we describe a new multi-module system at Brookhaven National Laboratory (BNL) in which an oxide molecular beam epitaxy (OMBE) is interconnected with an ARPES and a spectroscopic-imaging scanning tunneling microscopy (SI-STM) module. This new capability largely expands the range of complex-oxide materials and artificial heterostructures accessible to these two most powerful and complementary techniques for studies of electronic structure of materials. We also present the first experimental results obtained using this system — the ARPES studies of electronic band structure of a $La_{2-x}Sr_xCuO_4$ (LSCO) thin film grown by OMBE.

## 1 *Introduction*

Transition metal oxides host a variety of interesting electronic properties such as high-temperature superconductivity, colossal magneto-resistance, giant thermopower, piezo- and ferro-electricity, magnetism, topological states, etc. These cooperative electronic phenomena typically have characteristic temperatures of up to a few hundred degrees Kelvin, corresponding to a few tens of meV. ARPES and SI-STM are currently the two most powerful spectroscopic techniques that probe the electronic degree of freedom, in momentum and real space, respectively, at the relevant energy scale and with the required resolution. Being highly surface sensitive, however, both techniques require a flat and pristine sample surface. The latter can be prepared either by cleaving a single crystal *in situ* or by synthesizing a thin film in an interconnected adjacent chamber. Of these two techniques, the cleaving route presents much less of a technical challenge, and hence this has been prevalent so far. The drawback is that this greatly limits the range of available materials, basically to mechanically 'soft' layered oxides that can be cleaved and exfoliated.

In contrast, the *in-situ* integration scheme in principle can open to ARPES and/or STM a broad range of complex-oxide materials that can be synthesized using an OMBE with remarkable precision, control and reproducibility. Moreover, if an atomic-layer-by-layer[1] (ALL-MBE) synthesis mode is achieved, one can terminate the film growth at a desired atomic layer, as well as study the effects of epitaxial strain, proximity effects in heterostructures, etc. But this comes at the expense of substantial technical sophistication and challenges, and hence just a few systems in which complex-oxide MBE synthesis is integrated under ultra-high vacuum (UHV) conditions with ARPES[2,3,4,5,6,7,8] and/or STM[9], have been built and operated so far.

In what follows, we present a brief description of a newly constructed cluster system, OASIS — an acronym for **O**xide MBE, **A**RPES, and **SI**-**S**TM — at BNL, in which the three state-of-the-art components are linked via UHV interconnects. BNL has an extensive expertise with all three techniques. It also hosts the world's brightest third generation synchrotron, National Synchrotron Light Source (NSLS) II, located nearby. With that in mind, OASIS was designed in such a way that the OMBE-grown samples can be transferred using a portable UHV shuttle to several NSLS-II beamlines for further characterization using X-ray, infrared, or ultraviolet photons.

The OASIS facility spans two stories of the newly constructed Interdisciplinary Sciences building at BNL. OMBE and ARPES modules are installed in one laboratory on the first floor, while the STM is located in the underground ultra-low-vibration vault underneath. In the present article, we focus on the commissioned portion of the OASIS system, namely the OMBE and ARPES modules and their interconnection that allows for bidirectional sample transfer (see Fig. 1). The STM module is still in the commissioning stage, and we will briefly remark on some features relevant for its operation and integration with the other two OASIS modules.

The paper is organized as follows. We start this brief progress report with the description of the new sample holder in the section 2. OMBE and ARPES modules are described in the sections 3 and 4, respectively, while the section 5 explains how the two modules are interconnected for sample holder transfer. Representative *in-situ* ARPES data on OMBE-grown LSCO thin film are presented in section 6, and the section 7 concludes the article with a brief summary and outlook.

## **2** *Universal Sample Holder (USH)*

Interconnecting OMBE, ARPES and STM demanded designing an entirely new 'universal' sample holder (USH) that can be used in each of the three modules. Note that each of the three techniques have different and almost incompatible requirements. Oxide films are synthesized in OMBE at temperatures as high as 900 °C, under extremely corrosive environment of pure ozone. In order not to heat the surroundings, the sample holder should be as thermally decoupled from the heater-manipulator stage as possible while allowing access for reflection high-energy electron diffraction (RHEED) and atomic fluxes. In contrast, ARPES and STM are operated at cryogenic temperatures, which requires high electrical and thermal conductivity between the cold finger and the sample. The mechanical contact must also be completely rigid. Moreover, all the parts close to the sample must show negligible magnetism.

Combined, these requirements rule out the common structural refractory metals such as tungsten, molybdenum, or tantalum. Instead, for USH we use Inconel® alloy, which satisfies all the above requirements, but is expensive and hard to machine. The USH therefore should have the simplest geometry for ease of machining and re-usability. It also must be very small, so that the USH passes the sample transfer tube (OD less than 1") of the top-loading UHV STM cryostat. At the same time, it must enable handling using three completely different grabbers, enabling transfer in both horizontal (OMBE—ARPES) and vertical (OMBE—STM) directions, as well as continuous rotation with a well-controlled angle (ARPES).

We have designed such a USH (see Fig. 2A). The overall outer dimensions are 0.79" (D) × 0.67" (H), so it is small enough to fit the STM cryostat. It is composed of the top lid, the main body, and the bottom plate attached using four #0-80 screws. A substrate is clamped by the lid and the main body, while the bottom plate nut can be sacrificially removed in the case of screw seizing after extreme thermal cycling in corrosive environment. The USH has two outer grooves plus one inner groove, allowing it to be grabbed in several ways and directions, for maximal versatility. It is thus thermally and mechanically compatible with all three modules, OMBE, ARPES, and STM — i.e., it works smoothly with the respective receptacles and locking mechanisms and sample preparation stages including sputtering, electron-beam heating, and low-energy electron diffraction (LEED). The USH is hollow and enables infrared heating of the substrates from the back. The $10\times5\times1\text{mm}^3$ substrate geometry has been chosen in view of the dimensional constraints. Multiple additional modifications of the USH have been designed to enable a variety of additional treatments, measurements, and sample geometries.

This sample holder has been designed and verified to withstand the highly-oxidizing growth conditions of the OMBE chamber (the typical growth temperature of 600-700 °C and the pressure of $3\times10^{-5}$ Torr of ozone) without contaminating the film. It can be cooled down to liquid helium temperatures in STM and ARPES cryostats. The USH is compatible with STM requirements for maximum vibrational stability and largest possible contact surface with the receptacle for cooling the sample. It is also compatible with ARPES, which in addition to cooling, requires the sample holder to be non-magnetic to avoid the distortions of low-energy photoelectron paths. Additionally, the USH is compatible with the new beamline end-stations of the NSLS-II; the samples grown in OMBE can be studied in X-ray and ARPES systems using NSLS-II synchrotron radiation as the light source.

## 3 Oxide MBE

The synthesis capability of the OASIS is provided by an atomic layer-by-layer (ALL) ozone-assisted molecular beam epitaxy (OMBE) module. The system features the overall modular architecture[10], with the sources and analytical tools in autonomous UHV chambers outside of and separable from the main growth chamber, fully custom-designed and home-built specifically for the OASIS project.

The OMBE vacuum system is rather complex, consisting of over 20 UHV spaces in total. These include a two-stage load-lock (a fast entry turbo-pumped load-lock and a UHV ion-pumped storage buffer chamber) and a growth chamber (from UHV to $1\times10^{-4}$ Torr pressure of pure ozone). Each of the eight thermal effusion (Knudsen-, or K-) cells and eight quartz crystal monitors (QCMs) is housed in an individually valved UHV compartment. The vacuum in the chambers is achieved and maintained using a set of pumps (turbo, ion, titanium sublimation, non-evaporative getter) with a set of dry rough pumps housed in a separate and remote pump room. The growth chamber is pumped using a 1000 l/s magnetically levitated turbo-molecular pump (Seiko Seiki STP1003-CF10) and an additional ion pump (Varian StarCell 300) which is normally valved off during the growth. In addition to that, each of the 8 K-cell spools are equipped with an 80 l/s turbo-molecular pump (Pfeiffer HiPace 80) adding to the pumping speed of the main chamber during growth and providing differential pumping of ozone for each K-cell individually. The differential pumping effect is further enhanced by beam-collimating diaphragms resulting in an order of magnitude pressure drop between the main chamber and the K-cell. Moreover, the modular architecture allows for servicing the K-cells or QCMs independently, without venting the rest of the chamber or contaminating the other sources, thus minimizing the system downtime for recharging of the source material, maintenance, repairs, or upgrades.

The need to use the non-standard small sample holder (described in section 2, USH) impacted the design of the OMBE module. It precluded the use of a conventional commercial heater assembly; instead, a patented Thermo Riko infrared heater is used in combination with a custom sample manipulator developed in collaboration with Createc GmbH specifically for OASIS USH and sample transfer system. The heater has been tested to achieve temperatures above 900 °C in ozone.

The key of the heater design is that the radiation-generating infrared furnace is located outside of the UHV space (in air, allowing for easy replacement of the light bulb). The delivery of the radiation to the back of the sample is achieved using a quartz rod[11] via a water-cooled UHV sapphire viewport. The quartz rod remains cold and hence the sample holder and a receptacle that is supporting it are the only parts that are getting heated inside the chamber. Minimizing the number of heated surfaces improves the vacuum performance, eliminates potential problems of sample contamination, and, improves corrosion resistance of the heater in ozone. The temperature of the substrate is monitored using a pyrometer (Lumasense IGA 50-LO+) which can be operated in a proportional–integral–derivative (PID) loop with the heater and provides a laser spot for the alignment. Additionally, a camera equipped with a long-work-distance macro lens is mounted underneath the chamber to assess the surface and homogeneity of the sample temperature during growth and facilitate substrate alignment during transfers. A high degree of the substrate temperature homogeneity has been achieved which can be seen on figure 3A.

The OMBE module is equipped with a 35 keV RHEED system (STAIB Instruments GmbH). It allows for in-situ control of the growth process by monitoring RHEED patterns and the oscillations in the intensity of the electron beam diffracted from the sample surface. The RHEED system is also differentially pumped and the electron source is magnetically shielded. The electron source is equipped with both rocking and deflection degrees of freedom. One of the advantages of the contactless method of heating using infrared radiation is that it does not create electromagnetic interference with RHEED.

The manipulator of the system provides multiple degrees of freedom, including 2 tilts (along and perpendicular to RHEED axis), continuous rotation, and 3 degrees of translation, with the later 4 moved by stepper motors and controlled using either a wireless joystick or a computer.

In order to withstand highly corrosive growth conditions, all the parts within the hot zone of the heater (including the USH itself) are made out of Inconel. Additionally, to prevent degassing from the hot surfaces of the heater, the scattered radiation is shielded with a water-cooled spool surrounding the hot zone of the sample and the quartz rod.

The sample transfer system consists of a number of sample receptacles (mounted on revolvers attached to UHVDesign MagiDrive rotary feedthroughs) and a set of forks (mounted on UHVDesign magnetic manipulator arms) which interface with the USH. The system allows to introduce a sample from air via a quick access load-lock, pump down, degas, and then transfer and store up to 6 samples in a UHV buffer chamber. Each sample can then be transferred to the heater/manipulator stage located in the growth chamber for the film deposition. Once the growth is complete, the sample can be moved to the central distribution chamber of the OASIS system (grand central station; see section 5) for subsequent transfers into STM or ARPES modules. The sample transfer system is bidirectional, allowing to perform experiments in which the surface of the material can be prepared in OMBE, studied by ARPES, modified in the OMBE machine, moved back to ARPES, etc.

Since copper is hard to oxidize, in OMBE we use gaseous ozone ($O_3$) because of its extreme oxidizing power. To produce pure ozone, a commercial ozone generator (MKS AsTeX $O_3$MEGA AX8561-1021) is enhanced by adding a custom-designed cryogenic distillation system assembled within a gas safety cabinet. The gaseous ozone is introduced into the chamber by means of ½" OD SS316 tubing. Care has been taken to minimize the cracking of ozone within the gas handling system before the point of delivery by minimizing the number of valves and surfaces as well as shortening the path between the liquid phase and the valved and water-cooled injector of the growth chamber. The ultrapure gas handling system is built with all-stainless-steel and bakeable VCR-only components and features an independent turbo pump, a dry pump for ozone distillation (Pfeiffer ACP15) and multiple gauges.

In order to control the film stoichiometry, we use a set of 8 (one per K-cell) QCMs together with an elaborate (computerized, real-time) data-acquisition system (produced by Zensoft Inc. and dubbed ZenMBE). A network-based distributed automation system is implemented that allows for real-time control of shutters and various film recipes, control of interlocked operation of pneumatic gate valves and mechanical components of the sample transfer system, automated drift-compensated flux calibration with QCMs, real-time RHEED measurements, PID loops controlling ozone pressure, substrate heater, and K-cells, as well as monitoring of vacuum environment and logging of all support infrastructure peripherals.

# 4 ARPES

The ARPES system consists of an analysis chamber, a preparation chamber, a fast load-lock and a non-oxide MBE chamber. The analysis chamber is equipped with a noble-gas plasma discharge lamp, a hemispherical electron spectrometer, a Low Energy Electron Diffraction (LEED) spectrometer, a small service chamber and a 5-axis sample manipulator with the cryostat. It is a custom designed chamber, lined with double-layer Mu-metal® magnetic shield liners, perfectly mated with the double liners of the electron spectrometer that limit the residual magnetic field around the sample position for the ARPES studies to less than 0.1 µT. The chamber is pumped by turbo, ion and Ti-sublimation pumps, routinely achieving $5\times10^{-11}$ Torr of residual pressure. The electron spectrometer is a commercial Scienta R4000-WAL Electron Spectrometer with a wide acceptance angle (~30 deg), allowing simultaneous detection of a large window of the momentum space with ~0.1 deg precision. The best achievable energy resolution is of the order of 1 meV, sufficient to resolve fine spectral features responsible for very low energy/temperature phenomena in materials studied within the OASIS program. The operating kinetic energy range allows for a wide variety of possible photon sources for the excitation, including the low energy UV-lasers. The UV-photon source is the state of the art tabletop Scienta VUV5k microwave-driven plasma discharge lamp with a monochromator and a retractable small spot capillary. At He *I* radiation (21.22 eV), typically used for the excitation, the bandwidth is of the order of 1 meV, a perfect match to that of the electron analyzer. The small spot of the source (0.3 mm) allows studies of relatively small samples. The small spot will also allow a combinatorial analysis of films that will be synthesized in OMBE in the future, in which the composition will be quasi-continuously varied from one to the other side of the film. This will enable covering a large portion of the material's phase diagram with a single piece of sample grown under identical synthesis conditions. The sample manipulator is a fully motorized 5-axis manipulator (OmniVac HPM Cryo), with 3 translations and polar and tilt rotations, equipped with a Janis flow cryostat (ST-400-2) that allows cooling of the samples down to 6 K through a Cu braid. It is mounted on a short service chamber on top of the analysis chamber and can be valved off for maintenance. LEED spectrometer is OCI-Vacuum Engineering mini-LEED optics (model BDL450) for surface electron diffraction and Auger electron spectroscopy characterization of studied surfaces. The analysis chamber is also equipped with several alkali sources (SAES Getters), positioned so that the evaporation can be performed while the ARPES data are recorded, for *in-situ* modification of surface doping.

Attached to the main analysis chamber is a multi-purpose preparation chamber where the samples can be stored in UHV (base pressure $\sim 2\times 10^{-10}$ Torr) and additionally treated, if needed. Films synthesized in the OMBE are transferred there from the central distribution chamber (grand central station; see section 5). The preparation chamber is directly connected with the non-oxide MBE chamber and a fast load-lock entry to feed in non-oxide films and bulk samples, respectively. It is equipped with an ion-sputtering gun, several resistively heated and e-beam evaporators and the sample manipulator with the sample storage carousel for 12 samples and an e-beam heating stage where a sample can be heated up to 1200 °C.

The non-oxide MBE chamber has three resistively heated effusion cells and two e-beam sources for synthesis of transition-metal chalcogenides and other non-oxide films. Samples can be heated to 1500 °C via an e-beam heater. The thickness is monitored by a QCM. Integration of a RHEED is planned in the future.

## 5 *Film transfer*

OMBE / ARPES / STM modules are designed to exchange a USH via 'grand central station' (GCS) chamber (Fig. 1). The three units are located along mutually orthogonal directions relative to the GCS. In the coordinate system shown in Fig. 1, GCS defines the origin, while OMBE / ARPES / STM lie on positive x / y / z axes, respectively. Isolated from OMBE / ARPES chamber via gate valves, the GCS chamber can be independently evacuated with dedicated UHV infrastructure. The GCS chamber is mounted on an aluminum profile cage, which is firmly fixed on a machine base that is anchored to the concrete floor.

A USH is typically transferred from OMBE to GCS to ARPES chamber. When the film growth is completed, the USH is oriented such that the substrate points down. A platen-type fork at the end of a magnetic arm picks up the USH by prongs sliding into the outer groove farther from the substrate. Once the fork takes the USH out of the OMBE chamber, the fork and the USH are rotated by 90 deg before arrival at the GCS chamber so that the substrate normal is horizontal for the rest of the travel to the ARPES module. The lower prong of the fork is then guided by a matching slot in the stainless-steel block inside the GCS chamber (Fig. 2B, but fork should be vertical) to ensure that the cylindrical axis of USH is horizontal. Next, a claw-type coaxial grabber shown in Fig. 2B catches the USH using the outer groove closer to the substrate. In this grabber, a 'claw' can be closed or opened by a push-pull action of a collet on a two-shaft magnetic arm. After OMBE fork pulls back, the claw grabber loads the USH to a relay receptacle under the storage carousel of the ARPES preparation chamber. Another claw grabber subsequently feeds the USH into the ARPES main chamber, along a direction perpendicular to the GCS – ARPES claw grabber, for LEED and ARPES spectroscopy. The entire transfer process is fully reversible, thereby allowing for bi-directional USH transfer. Note that a fork can change polar orientation of the USH, while claw grabber can change the azimuthal orientation. ARPES module needs the latter functionality to load the sample in an arbitrary azimuthal angle. For transfer to STM, the sample remains pointing down as depicted in Fig. 2B.

The OMBE / GCS / ARPES chambers are approximately 4ft away from their nearest neighbors. Large inter-chamber distances provide ample space for convenient operation and maintenance access. The effects of chamber misalignment and magnetic arm shaft drooping upon full stretch are minimized by using several custom designed guiding pieces - tapered plastic (PEEK) sleeves assisting shaft alignment, slotted block inside GCS chamber guiding the OMBE fork, and standard port aligners are a few examples.

Since the distance scale is rather large, and since there are complex gadgets present in between the chambers, real-time naked-eye monitoring of USH movement through chamber viewports while manipulating magnetic arms is impractical. We developed a system of Raspberry Pi cameras and LCD monitors for remote monitoring. As most of the sample transfer steps are motorized and remotely controlled, the operator can stay next to the monitors during the entire transfer.

## 6 Results

The 10-unit cells thick LSCO film was grown on LaSrAlO$_4$ substrate epi-polished perpendicular to the [001] crystallographic direction following 2-layer thick overdoped (x = 0.4) and Cu-rich (Cu1.1 stoichiometry) buffer layer. The USH carrying the clamped substrate was loaded through the OMBE load-lock chamber. The films were grown by depositing La, Sr and Cu, sequentially shuttered to form full LSCO layers one at a time. The ozone partial pressure was kept constant during film growth at $p(O_3) = 3\times10^{-5}$ Torr growth chamber pressure (while the local pressure at the injector was about 100mTorr and the detected 48 amu peak reaching $1\times10^{-6}$ Torr on a residual gas analyzer not directly exposed to the ozone flux). The substrate temperature was fixed at $T_s$ = 650°C measured with an optical pyrometer (set at the emissivity ε = 1). Typical RHEED pattern during growth shown in Fig. 3B demonstrates high crystalline quality of the film, comparable to those grown by one of us (I. B.)[12,13,14]. Once the film growth was complete, the substrate was slowly (over 1 hour) cooled down to room temperature in high-pressure ($3\times10^{-5}$ Torr) ozone atmosphere to prevent loss of oxygen from the film. After that, the ozone was evacuated as quickly as possible with additional pumping provided by an ion pump and the film was transported to the ARPES chamber. During the transfer, which took less than 5 min to complete, the total pressure was below $5\times10^{-9}$ Torr in the OMBE growth chamber and UHV along the way. The sharp LEED pattern in Fig. 3C acquired at 122eV incident electron beam prior to ARPES measurement exhibits tetragonal structure indicating high crystallinity. For high-resolution ARPES experiments, monochromatized He *II* radiation ($hv$ = 40.8 eV) was used for the excitation. The energy resolution was set by selecting the entrance slit size and the pass energy of the hemispherical analyzer to ~10 meV, while the angular resolution was ~0.2 deg. Temperature was measured using a calibrated Si-diode sensor mounted on the USH receptacle. The Fermi level was pre-determined by measuring a metallic sample at all possible combinations of slit widths and pass energies used in the study.

Fig. 3D represents the ARPES intensity at the Fermi level, integrated over the interval of ± 8 meV, for one of the first successfully OMBE-grown LSCO films, showing a sharp Fermi surface (FS). Spectra along the line cuts roughly parallel to $\Gamma$-$\bar{M}$ direction of the Brillouin zone in Fig. 3E exhibit sharp electronic states, dispersing in a way so that they form the electron-like Fermi surface with a narrow energy and momentum linewidth, comparable with the ARPES data from published studies on cleaved LSCO single crystals[15,16,17,18,19,20,21,22,23]. The FS encloses an electron pocket around the zone center, therefore clearly indicating strongly overdoped LSCO film. Solid curves in Fig. 3D are the tight-binding fit of the measured FS contour[24]. The enclosed FS area indicates that the hole doping level is significantly higher than the nominal doping of x = 0.16. The strong overdoping inferred from the FS area is consistent with the fact that the spectral gap was not observed down to the lowest temperature of less than 10 K, and consecutive mutual inductance measurements showing no superconductivity down to 4 K. The surface was extremely stable, producing nearly identical intensity map a day after the initial measurement with the sample kept in the ARPES analysis chamber.

## 7 Outlook

In summary, state-of-the-art OMBE and ARPES units are now successfully integrated, owing to the new 'universal' sample holder and the intermediary vacuum hardware for connecting the two systems. LSCO films were synthesized in the OMBE chamber and transferred to the ARPES module, where their electronic structure was studied. The sharp diffraction patterns in LEED and sharp electronic states forming a well-defined Fermi surface are comparable with the published data from bulk LSCO crystals, thus indicating that the films are of exceptionally high quality.

These highly promising results from the initial test runs attest to the potential power of the OASIS system for spectroscopic study of complex oxide films. In another early experiment, bi-directional traffic of USH between the OMBE and the ARPES modules, combined with exceptional oxidizing power of ozone in the OMBE, enabled us to reproducibly reach extremely high doping levels in Bismuth-based cuprate, hitherto inaccessible with conventional oxygen-based annealing.

In the near future, additional components to be added to the existing setup are expected to provide further analytical capabilities. The integration and commissioning of SI-STM, the critically important next component of the OASIS system, will be completed soon. Other planned extensions include a module for an *in-situ* micro four-point transport and in-situ mutual inductance (magnetic susceptibility) measurements. Also, the planned upgrade of OMBE with the combinatorial synthesis will enable studies of a wide composition regime of a material of interest on a single 10×5 mm$^2$ sample. OASIS will thus open the path to a variety of electron-spectroscopy experiments that were not feasible so far.

*Acknowledgement*

This research was supported by the U.S. Department of Energy, Office of Basic Energy Sciences, under Contract No. DE-SC0012704. I.K.D. acknowledges the technical support of Robert J. Sundling and the generous financial support of the BNL Gertrude and Maurice Goldhaber Distinguished Fellowship.

**FIGURE CAPTIONS**

Figure 1: **Interconnected OMBE – ARPES modules**

A. 3D CAD model showing the overall layout of the interconnected OMBE – ARPES module via GCS chamber. The square floor tiles are 2ft×2ft wide each.

B. A picture of the actual system, corresponding to the CAD model in A.

Figure 2: **Universal sample holder (USH) and grand central station (GCS)**

A. 3D CAD model of universal sample holder. A USH is composed of top lid (light blue) – main body (light brown) – plate nut (dark blue) stacked vertically. A substrate (a rectangular block in cyan) is pressed down by the four corner pads of the top lid. Two parallel circular grooves are for fork grabbers, while the straight vertical slot is for locking loading the USH into the STM.

B. 3D CAD model of the grabber guide assembly located inside the GCS chamber, along with three different types of grabbers optimized for each module. The stainless-steel block in yellow has four horizontal slots to support OMBE fork prongs while STM grabber picks up the USH (as shown in this figure), and two vertical slots for USH transfer to ARPES claw-grabber. The PEEK plastic piece (in pink) on top of the block guides the STM grabber using a cylindrical cutout at the center.

Figure 3: **OMBE-grown LSCO film characterized by ARPES**

A. An image of a USH and a substrate heated in the OMBE chamber. The measured temperature of the substrate (orange rectangle at the center) is approximately 700 °C. The spatial homogeneity of the orange color indicates the remarkable uniformity of the substrate temperature.

B. A representative RHEED pattern after deposition of 12 layers of LSCO. The specular spot, "streaky" RHEED pattern with "sidebands", and the presence of Laue circles and Kikuchi lines indicate high quality of thin film without secondary phase precipitates. The RHEED pattern with the 6×6 reconstruction is characteristic of the overdoped LSCO sample.

C. LEED pattern of the LSCO film measured with 122 eV incident electron beam, indicative of the high crystallinity of the synthesized film.

D. ARPES intensity map of the LSCO film, integrated over $E_F \pm 8$ meV range. The solid curve represents tight-binding fit to the mapped Fermi surface. The doping level is estimated to be $p = 0.275$ and $T_c$ is lower than 4.5 K.

E. Energy-momentum dispersion plots along the dashed lines in D.

Figure 1

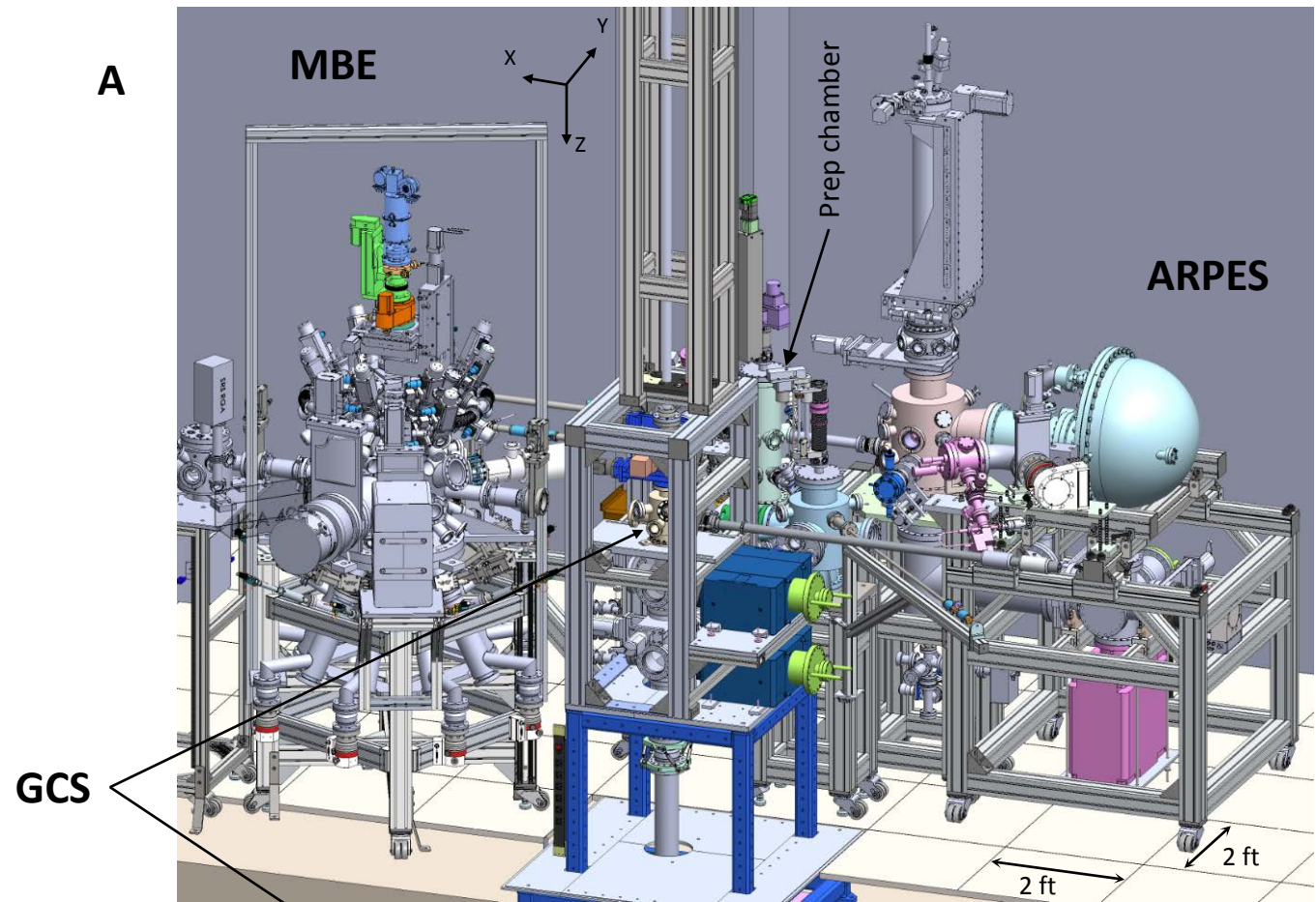

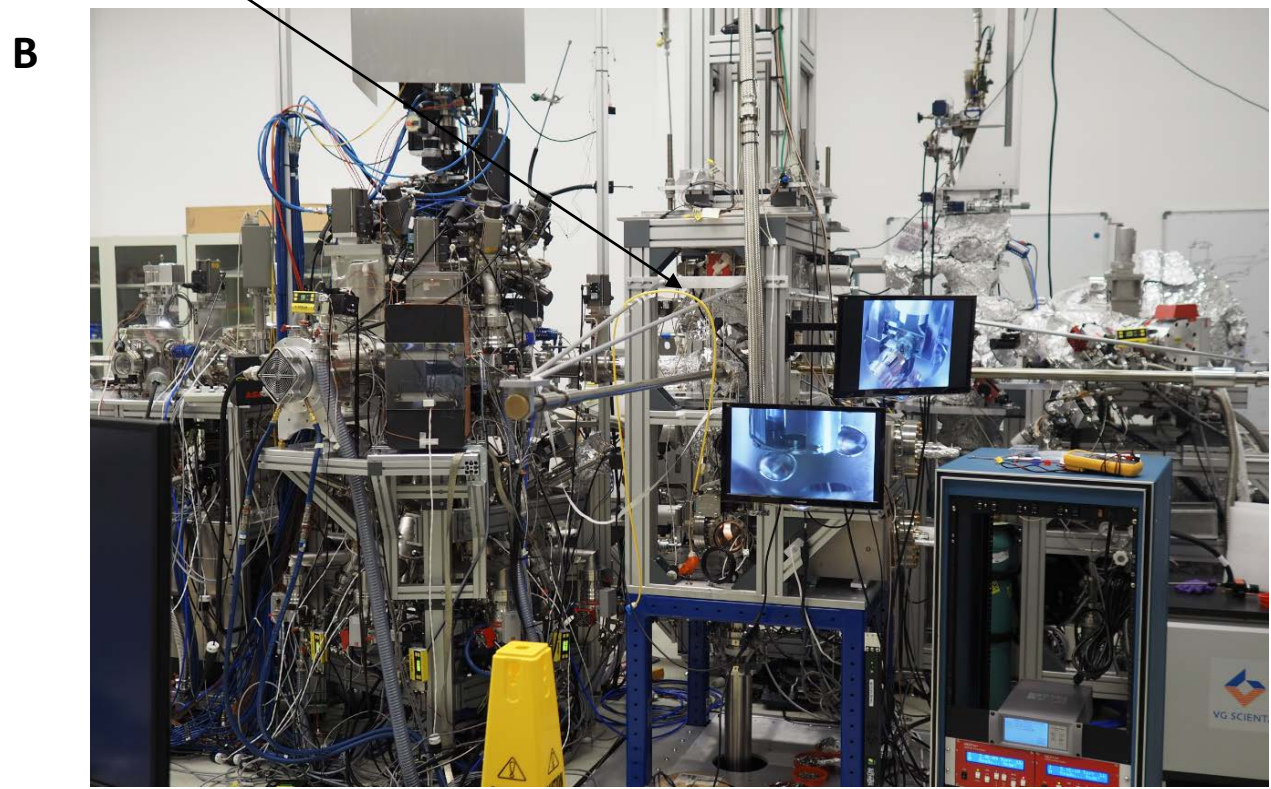

Figure 2

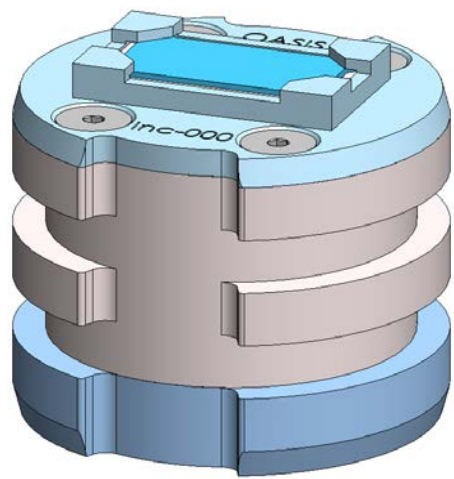

A

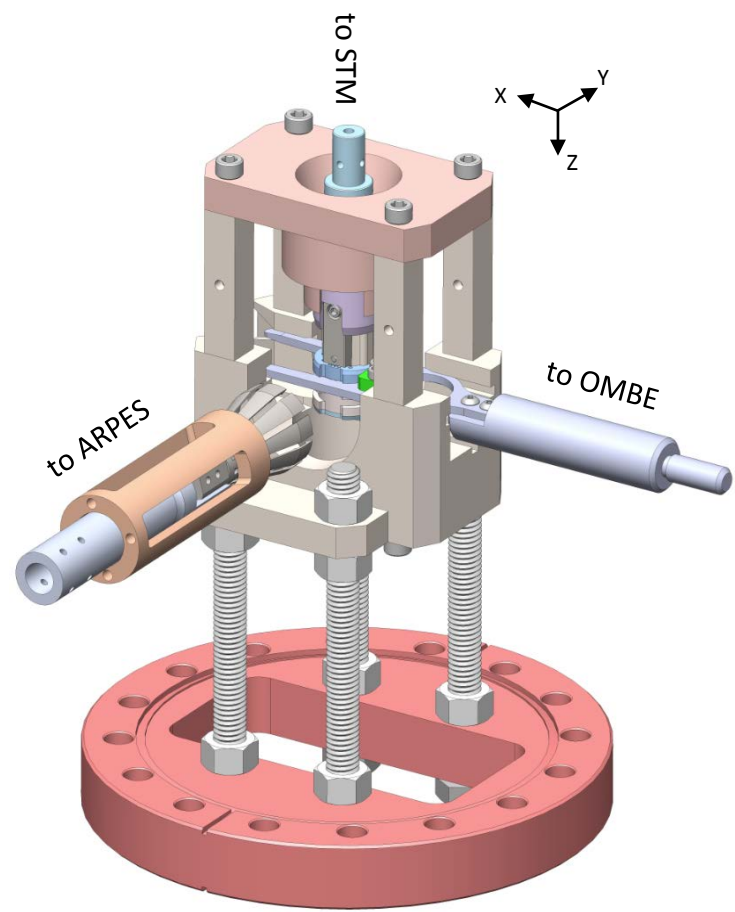

B

Figure 3

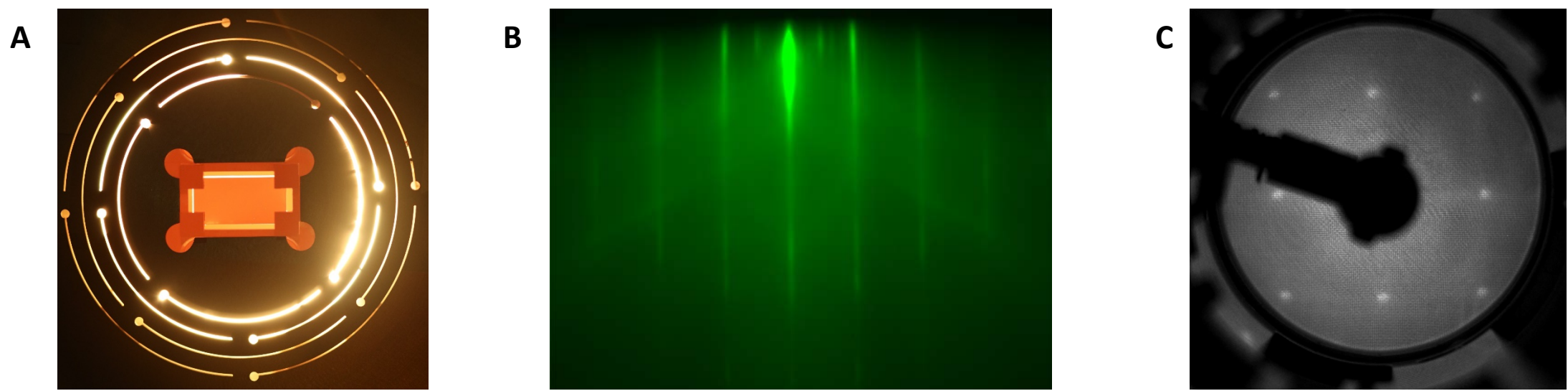

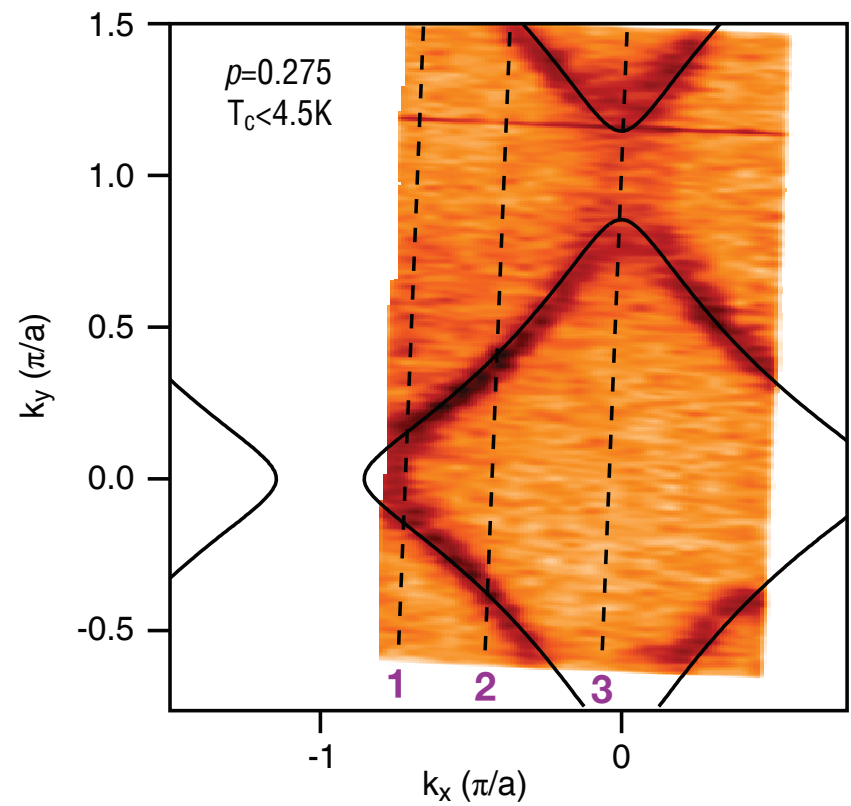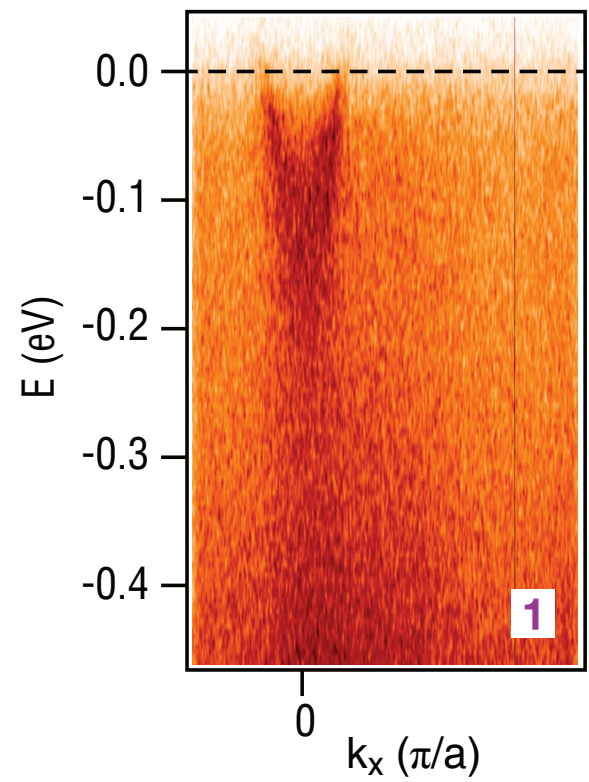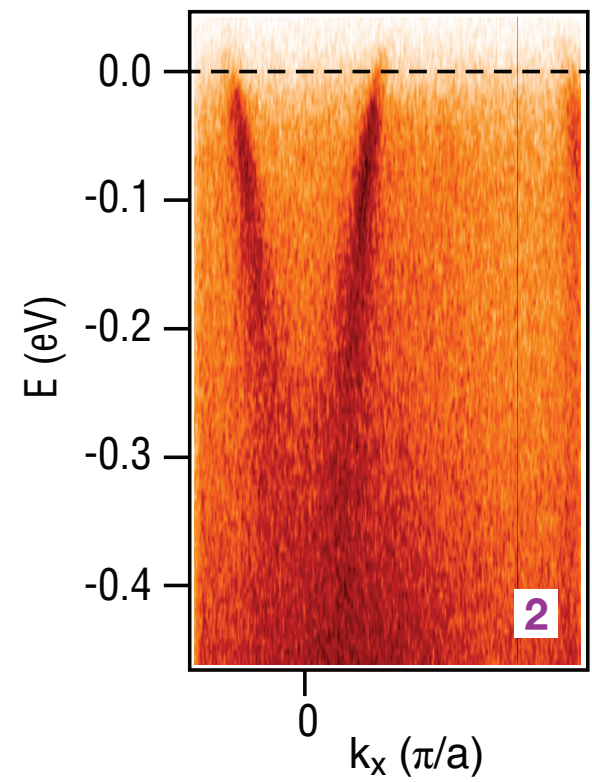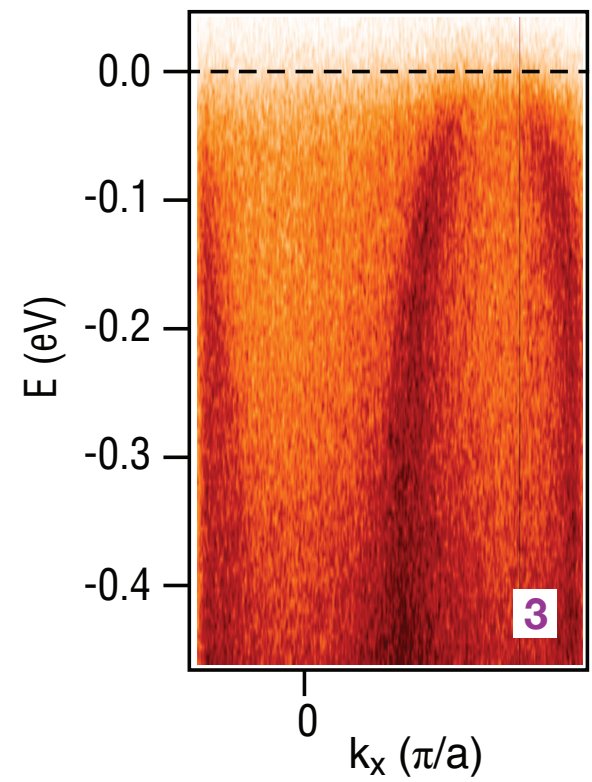